\documentclass[twocolumn,trackchanges]{aastex701}
\usepackage{xspace}
\usepackage{amsmath}
\usepackage{subcaption}

\usepackage{comment}

\usepackage{graphicx}
\usepackage{blindtext}

\newcommand{\ixpe}{\textit{IXPE}\xspace}

\newcommand{\ixpeobssim}{\textsc{ixpeobssim}\xspace}
\newcommand{\axj}{AX~J1745.6$-$2901\xspace}
\newcommand{\maxij}{MAXI~J1744$-$294\xspace}

\shorttitle{X-ray polarimetry of AX J1745.6--2901 with \ixpe}
\shortauthors{Miku\v{s}incov\'{a} et al.}

\begin{document}

\title{The First X-Ray Polarimetry of an Eclipsing Low-Mass X-Ray Binary: \\ Serendipitous IXPE Observation of AX J1745.6--2901}

\author[0000-0001-7374-843X]{Romana Miku\v{s}incov\'{a}}
\affiliation{INAF Istituto di Astrofisica e Planetologia Spaziali, Via del Fosso del Cavaliere 100, 00133 Roma, Italy}
\email{romana.mikusincova@inaf.it}

\author[0009-0001-4644-194X]{Lorenzo Marra}
\affiliation{INAF Istituto di Astrofisica e Planetologia Spaziali, Via del Fosso del Cavaliere 100, 00133 Roma, Italy}
\email{lorenzo.marra@inaf.it}

\author[0000-0001-9404-1601]{Hemanth Manikantan}
\affiliation{INAF Istituto di Astrofisica e Planetologia Spaziali, Via del Fosso del Cavaliere 100, 00133 Roma, Italy}
\email{hemanth.manikantan@inaf.it}

\author[0000-0002-4622-4240]{Stefano Bianchi}
\affiliation{Dipartimento di Matematica e Fisica, Universit\`a degli Studi Roma Tre, Via della Vasca Navale 84, 00146 Roma, Italy}
\email{stefano.bianchi@uniroma3.it}

\author[0000-0002-6384-3027]{Fiamma Capitanio}
\affiliation{INAF Istituto di Astrofisica e Planetologia Spaziali, Via del Fosso del Cavaliere 100, 00133 Roma, Italy}
\email{fiamma.capitanio@inaf.it}

\author[0000-0003-4872-8159]{Sudip Chakraborty}
\affiliation{Science and Technology Institute, Universities Space and Research Association, Huntsville, AL 35805, USA}
\email{schakraborty2@usra.edu}

\author[0009-0009-3289-3767]{Raul Ciancarella}
\affiliation{Independent Researcher}
\email{raul.cianca@gmail.com} 

\author[0000-0003-4925-8523]{Enrico Costa}
\affiliation{INAF Istituto di Astrofisica e Planetologia Spaziali, Via del Fosso del Cavaliere 100, 00133 Roma, Italy}
\email{enrico.costa@inaf.it} 

\author[0000-0002-2498-0213]{Nicolas {De Angelis}}
\affiliation{INAF Istituto di Astrofisica e Planetologia Spaziali, Via del Fosso del Cavaliere 100, 00133 Roma, Italy}
\email{nicolas.deangelis@inaf.it}

\author[0000-0002-1793-1050]{Melania Del Santo}
\affiliation{INAF Istituto di Astrofisica Spaziale e Fisica Cosmica, Via U. La Malfa 153, I-90146 Palermo, Italy}
\email{melania.delsanto@inaf.it}

\author[0000-0003-1533-0283]{Sergio Fabiani}
\affiliation{INAF Istituto di Astrofisica e Planetologia Spaziali, Via del Fosso del Cavaliere 100, 00133 Roma, Italy}
\email{sergio.fabiani@inaf.it} 

\author[0000-0003-1074-8605]{Riccardo Ferrazzoli}
\affiliation{INAF Istituto di Astrofisica e Planetologia Spaziali, Via del Fosso del Cavaliere 100, 00133 Roma, Italy}
\email{riccardo.ferrazzoli@inaf.it}

\author[0000-0002-9719-8740]{Vittoria E. Gianolli}
\affiliation{Department of Physics and Astronomy, Clemson University, Clemson, SC, 29634, USA}
\email{vgianol@clemson.edu}

\author[0000-0002-0642-1135]{Andrea Gnarini}
\affiliation{Dipartimento di Matematica e Fisica, Universit\`a degli Studi Roma Tre, Via della Vasca Navale 84, 00146 Roma, Italy}

\email{andrea.gnarini@uniroma3.it}

\author[0000-0002-5311-9078]{Adam Ingram}
\affiliation{School of Mathematics, Statistics, and Physics, Newcastle University, Newcastle upon Tyne NE1 7RU, UK}
\email{adam.ingram@newcastle.ac.uk}

\author[0000-0002-6126-7409]{Shifra Mandel}
\affiliation{Columbia Astrophysics Laboratory, Columbia University, New York, NY 10027, USA}
\email{ss5018@columbia.edu}

\author[0000-0003-4216-7936]{Guglielmo Mastroserio}
\affiliation{Scuola Universitaria Superiore IUSS Pavia, Palazzo del Broletto, piazza della Vittoria 15, I-27100 Pavia, Italy}
\email{guglielmo.mastroserio@iusspavia.it}

\author[0000-0002-2152-0916]{Giorgio Matt}
\affiliation{Dipartimento di Matematica e Fisica, Universit\`a degli Studi Roma Tre, Via della Vasca Navale 84, 00146 Roma, Italy}
\email{giorgio.matt@uniroma3.it}

\author[0000-0002-9709-5389]{Kaya Mori}
\affiliation{Columbia Astrophysics Laboratory, Columbia University, New York, NY 10027, USA}
\email{kaya@astro.columbia.edu}

\author[0000-0003-3331-3794]{Fabio Muleri}
\affiliation{INAF Istituto di Astrofisica e Planetologia Spaziali, Via del Fosso del Cavaliere 100, 00133 Roma, Italy}
\email{fabio.muleri@inaf.it}

\author[0009-0004-5622-1854]{Simone Pagliarella}
\affiliation{INAF Istituto di Astrofisica e Planetologia Spaziali, Via del Fosso del Cavaliere 100, 00133 Roma, Italy}
\affiliation{Tor Vergata University of Rome, Via Della Ricerca Scientifica 1, 00133 Roma, Italy}
\affiliation{Dipartimento di Fisica, Università degli Studi di Roma `La Sapienza', P.le Aldo Moro 2, 00133 Roma, Italy}
\email{simone.pagliarella@inaf.it}

\author[0009-0003-8610-853X]{Maxime Parra}
\affiliation{Department of Physics, Ehime University, 2-5, Bunkyocho, Matsuyama, Ehime 790-8577, Japan}
\email{maxime.parrastro@gmail.com} 

\author[0000-0001-6061-3480]{P.O. Petrucci}
\affiliation{Universit\'e Grenoble Alpes, CNRS, IPAG, 38000 Grenoble, France}
\email{pierre-olivier.petrucci@univ-grenoble-alpes.fr}

\author[0000-0001-5418-291X]{Jakub Podgorn{\'y}}
\affiliation{Astronomical Institute of the Czech Academy of Sciences,
Bo\v{c}n\'{i} II 1401/1, 14100 Praha 4, Czech Republic}
\email{jakub.podgorny@asu.cas.cz}

\author[0000-0002-0983-0049]{Juri Poutanen}
\affiliation{Department of Physics and Astronomy,  20014 University of Turku, Finland}
\email{juri.poutanen@gmail.com}

\author[0000-0002-2381-4184]{Swati Ravi}
\affiliation{MIT Kavli Institute for Astrophysics and Space Research, Massachusetts Institute of Technology, Cambridge, MA 02139, USA}
\email{swatir@mit.edu}

\author[0000-0002-7781-4104]{Paolo Soffitta}
\affiliation{INAF Istituto di Astrofisica e Planetologia Spaziali, Via del Fosso del Cavaliere 100, 00133 Roma, Italy}
\email{paolo.soffitta@inaf.it} 

\author[0000-0002-5872-6061]{James~F.~Steiner}
\affiliation{Center for Astrophysics, Harvard \& Smithsonian, Cambridge, MA 02138, USA}
\email{james.steiner@cfa.harvard.edu}

\author[0009-0007-0537-9805]{Antonella Tarana}
\affiliation{INAF Istituto di Astrofisica e Planetologia Spaziali, Via del Fosso del Cavaliere 100, 00133 Roma, Italy}
\email{Antonella.tarana@inaf.it}

\author[0000-0002-1768-618X]{Roberto Taverna}
\affiliation{Dipartimento di Fisica e Astronomia, Universit\`{a} degli Studi di Padova, Via Marzolo 8, 35131, Padova, Italy}
\email{roberto.taverna@unipd.it}

\author[0000-0002-2055-4946]{Francesco Ursini}
\affiliation{Dipartimento di Matematica e Fisica, Universit\`a degli Studi Roma Tre, Via della Vasca Navale 84, 00146 Roma, Italy}
\email{francesco.ursini@uniroma3.it} 

\author[0000-0002-5767-7253]{Alexandra Veledina}
\affiliation{Department of Physics and Astronomy, 20014 University of Turku, Finland}
\affiliation{Nordita, KTH Royal Institute of Technology and Stockholm University, Hannes Alfv\'ens v\"ag 12, SE-10691 Stockholm, Sweden}
\email{alexandra.veledina@gmail.com}

\author[0000-0002-1481-1870]{Federico M. Vincentelli}
\affiliation{INAF Istituto di Astrofisica e Planetologia Spaziali, Via del Fosso del Cavaliere 100, 00133 Roma, Italy}
\affiliation{School of Physics and Astronomy, University of Southampton, University Road, Southampton SO17 1BJ, UK}
\email{vincentelli.astro@gmail.com}

\author[0000-0003-1133-1684]{Anastasiya Yilmaz}
\affiliation{INAF Istituto di Astrofisica e Planetologia Spaziali, Via del Fosso del Cavaliere 100, 00133 Roma, Italy}
\email{anastasiya.yilmaz@inaf.it}

\author[0000-0003-2743-6632]{Barbara De Marco}
\affiliation{Departament de Fis\'{i}ca, EEBE, Universitat Polit\`ecnica de Catalunya, Av. Eduard Maristany 16, S-08019 Barcelona, Spain}
\email{barbara.de.marco@upc.edu}

\author[0000-0003-0976-8932]{Maitrayee Gupta}
\affiliation{Astronomical Institute of the Czech Academy of Sciences, Bo\v{c}n\'{i} II 1401/1, 14100 Praha 4, Czech Republic}
\email{maitrayee.gupta@asu.cas.cz}

\author[0000-0001-6894-871X]{Vladislav Loktev}
\affiliation{School of Mathematics, Statistics, and Physics, Newcastle University, Newcastle upon Tyne NE1 7RU, UK}
\email{loktev.astro@gmail.com}

\author[0000-0002-7930-2276]{Thomas D. Russell}
\affiliation{INAF Istituto di Astrofisica Spaziale e Fisica Cosmica, Via U. La Malfa 153, I-90146 Palermo, Italy}
\email{thomas.russell@inaf.it}

\author[0000-0003-2931-0742]{Ji\v{r}\'{i} Svoboda}
\affiliation{Astronomical Institute of the Czech Academy of Sciences, Bo\v{c}n\'{i} II 1401/1, 14100 Praha 4, Czech Republic}
\email{jiri.svoboda@asu.cas.cz}

\author[0000-0002-6562-8654]{Francesco Tombesi}
\affiliation{Tor Vergata University of Rome, Via Della Ricerca Scientifica 1, 00133 Roma, Italy}
\email{francesco.tombesi@roma2.infn.it} 

\author[0000-0002-2967-790X]{Shuo Zhang}
\affiliation{Department of Physics and Astronomy, Michigan State University, East Lansing, MI 48824, USA}
\email{zhan2214@msu.edu}


\begin{abstract}
We present the first X-ray polarimetric measurement of the neutron star low-mass X-ray binary system \axj conducted by the Imaging X-ray Polarimetry Explorer (\ixpe) satellite. This transient source, located within $ \sim $1\farcm5 of the Galactic center, was observed serendipitously during a \maxij observation with a duration of 150~ks. The complex nature of the region in which \axj is located poses a challenge for studying its polarization. By performing a detailed analysis of the contamination from \maxij and the Galactic center diffuse emission, we find the source polarization degree PD~=~14.7\%~$\pm$~4.0\% and polarization angle PA~=~122\degr~$\pm$~8\degr. The phase-resolved analysis shows increase in polarization during the eclipse phase, with PD~=~34.2\%~$\pm$~8.7\%, suggesting that the polarization-inducing mechanisms are of scattering nature, probably originating from disk winds. 
\end{abstract}



\section{Introduction} 

Neutron star low-mass X-ray binaries (NS-LMXBs) are systems in which a neutron star accretes material from a low-mass companion star, typically via Roche-lobe overflow and the subsequent formation of an accretion disk \citep{Bhattacharya1991, vanderKlis2006}. These binaries are among the most luminous X-ray sources in the Galaxy, and their observable properties are shaped by a combination of accretion physics, magnetic field interactions, and geometrical orientation \citep{Psaltis2008, Nielsen2023}. NS-LMXBs can appear either as persistent emitters or as transients, depending on the long-term stability of the accretion disk \citep{Tanaka1996, King1998}. Transient systems undergo episodic outbursts, during which the X-ray luminosity can increase by several orders of magnitude before returning to a quiescent state. These outbursts are generally understood to be the result of thermal-viscous instabilities in the accretion disk \citep{King+96, Lasota2001}, and offer key opportunities to study accretion physics across a wide range of mass inflow rates \citep{Done2007}.

Throughout an outburst, transient NS-LMXBs show complex broadband spectra: thermal X-ray emission from the disk and neutron star surface, hard X-ray components simply modeled with a power-law resulting from Comptonization by thermal electrons in the corona, and often, relativistic jets and/or winds \citep{Migliari2006, Church2006, Nielsen2023}. Their spectra evolve through well-defined states \citep{Hasinger1989, vanderKlis1994}, and timing analysis often reveal coherent pulsations, quasi-periodic oscillations (QPOs), and type-I thermonuclear X-ray bursts \citep{Strohmayer2006, vanderKlis2006}. Systems with high orbital inclination may also display eclipses and dipping behavior, providing direct constraints on the geometry of the accretion flow \citep{White1982, Hyodo2009}. These characteristics make transient NS-LMXBs powerful laboratories for studying accretion physics, boundary-layer (BL, region where the accretion disk meets the neutron star) emission, and the coupling between accretion state and outflows \citep{Inogamov1999, Popham_2001}. 
Despite extensive spectral and timing studies, many aspects of their geometry and radiative processes remain difficult to probe, particularly those involving scattering and asymmetries in the disk structure \citep{Done2007}. X-ray polarimetry provides a promising new tool to directly probe these geometric and radiation characteristics by offering access to emission geometry, scattering environments, and magnetic field configurations \citep{Schnittman+09, Churazov2017}.
Through model-dependent spectro-polarimetry, the polarization degree (PD) of the BL of a few Z-sources has been measured: in the case of GX~5$-$1, \cite{Fabiani2024} identified polarization produced by the boundary layer at values 5.7\%$\pm$1.4\% while the source was in the horizontal branch (HB) and 4.3\%$\pm$2.0\% integrated over normal (NB) and flaring branches (FB); a PD of the BL component $ \approx 4$\% for Cygnus X-2 possibly in the normal branch was reported by \cite{Farinelli2023}; in the case of XTE~J1701$-$462 \citep{Cocchi2023} a value of 6.0\%$\pm$0.5\% was found for the hard blackbody component while the source was in the HB, with a drop of PD to 1.2\%$\pm$0.7\% in the NB. Interesting results have been achieved in the Atoll sources as well. The reflection component present in the spectra of GX~9+9 \citep{Ursini2023} was found to be polarized at 10\%, resulting in 3$\pm$1\% polarization of BL. In the case of GS~1826$-$238, only an upper limit on PD was found at $\lesssim 1.3\%$ \citep{Capitanio2023}. On the other hand, \cite{DiMArco2023} have found 4U~1820$-$303 to have the reflection component polarized at 5.3$\pm$0.2\%.

\axj~is a transient NS-LMXB located just $\sim1\farcm5$  from Sgr A$^{*} $, at the dynamical center of the Milky Way \citep{Maeda1996, Kennea1996, Hyodo2009, Ponti2015}. Its location in the densely populated and strongly absorbed Galactic Center region \citep{Morris1996, Muno2009} poses observational challenges. 
The nature of the compact object is confirmed by the detection of type-I X-ray bursts \citep{Maeda1996, Hyodo2009}.
The source exhibits periodic eclipses every $\sim$8.35 hours together with complex dipping behavior -- signatures of a nearly edge-on disk geometry \citep{Maeda1996, Hyodo2009}. The estimate on the source inclination is $ i \sim 70\degr$--$80\degr$ \citep{Ponti2015, Ponti2018}.
Studying this source provides a valuable opportunity to probe accretion and outflow physics in a high-inclination system \citep{Hyodo2009, Ponti2015, Ponti2018}.

The system shows strong X-ray variability, transitioning between long, bright outbursts and deep quiescence \citep{Chen1997, Ponti2015}. Archival XMM–Newton observations reveal that the source remains in outburst roughly one-third of the time, where it alternates erratically between hard and soft spectral states, with 3--10 keV fluxes of (1--5)$ \times 10^{-11}\, \rm{erg\,cm^{-2}\,s^{-1}} $ in the former, and (1--3)$\times 10^{-10}\, \rm{erg\,cm^{-2}\, s^{-1}}$ in the latter \citep{Ponti2018}. In the soft state, Fe~\textsc{xxv} and Fe~\textsc{xxvi} absorption lines are prominent, consistent with the presence of a highly ionized wind \citep{Ponti2015, Nielsen2023}. These features disappear in the hard state \citep{Ponti2018}, but this cannot be linked to the evolution of the outflow itself, as thermal instabilities prevent the formation of absorption lines from highly ionized elements when the plasma is illuminated by radiation with a hard spectrum \citep{Frank2002, Bianchi2017}. The absorbing gas has a large column density $ (N_{\rm H} \geq 10^{23}\, \rm{cm^{-2}})$ and is turbulent $ (v_{\rm turb} \geq 500\, \rm{km\,s^{-1}}) $, with small  K${\alpha}$/K${\beta}$ line ratios for highly ionized iron lines further indicating large column depths \citep[Matsunaga et al., 
in prep.;][]{Proga_2004, Ponti2015}.

The soft-state spectrum is well modeled by thermal components --- typically a disk blackbody and a blackbody associated with cooling radiation from the NS surface --- along with interstellar dust scattering and iron emission lines \citep{Mitsuda1984, Makishima1986, Hyodo2009}. A Comptonized disk, and thus traces of corona, are also found to fit the soft spectral state well \citep{Ponti2018}, although a thermal-Comptonization component would be weaker than the others. This makes \axj particularly interesting to verify the accretion disk polarization models \citep{Chandrasekhar+60, Loskutov1982}.

High-inclination LMXBs like \axj offer an especially favorable viewing angle to the system geometry for X-ray polarimetric studies, which can probe disk symmetry, constrain the orientation of reprocessing structures, and reveal the presence of scattering media such as winds or bulges \citep{Krawczynski2011, Fabiani2014a, Fabiani2014b, Churazov2017, Nitindala2025}. In particular, asymmetries of the disk \citep{Jin2018}, boundary or spreading layer, and absorbing material are expected to produce measurable polarization signals \citep{Schnittman+09, Farinelli2024, Bobrikova2025}. The absence of a strong Comptonized corona in \axj further enhances the interpretability of any detected polarization, making it an ideal candidate for probing the origin and structure of the component responsible for the soft-state emission \citep{Nielsen2023}.

In this paper, we report on the first X-ray polarimetric observation of the transient source \axj by the Imaging X-ray Polarimetry Explorer \citep[IXPE;][]{Weisskopf+22, Soffitta2021}. We present the observed energy-averaged and time-resolved polarimetric properties of the source. The data reduction is described in Section~\ref{sec:methods}, and the results of data analysis and their implications for the properties of the system is presented in Section~\ref{sec:results}. We discuss the analysis and the interpretation of the data in Section~\ref{sec:discussion} and finally, we summarize our findings in Section~\ref{sec:conclusions}.

\begin{figure*} 

\plotone{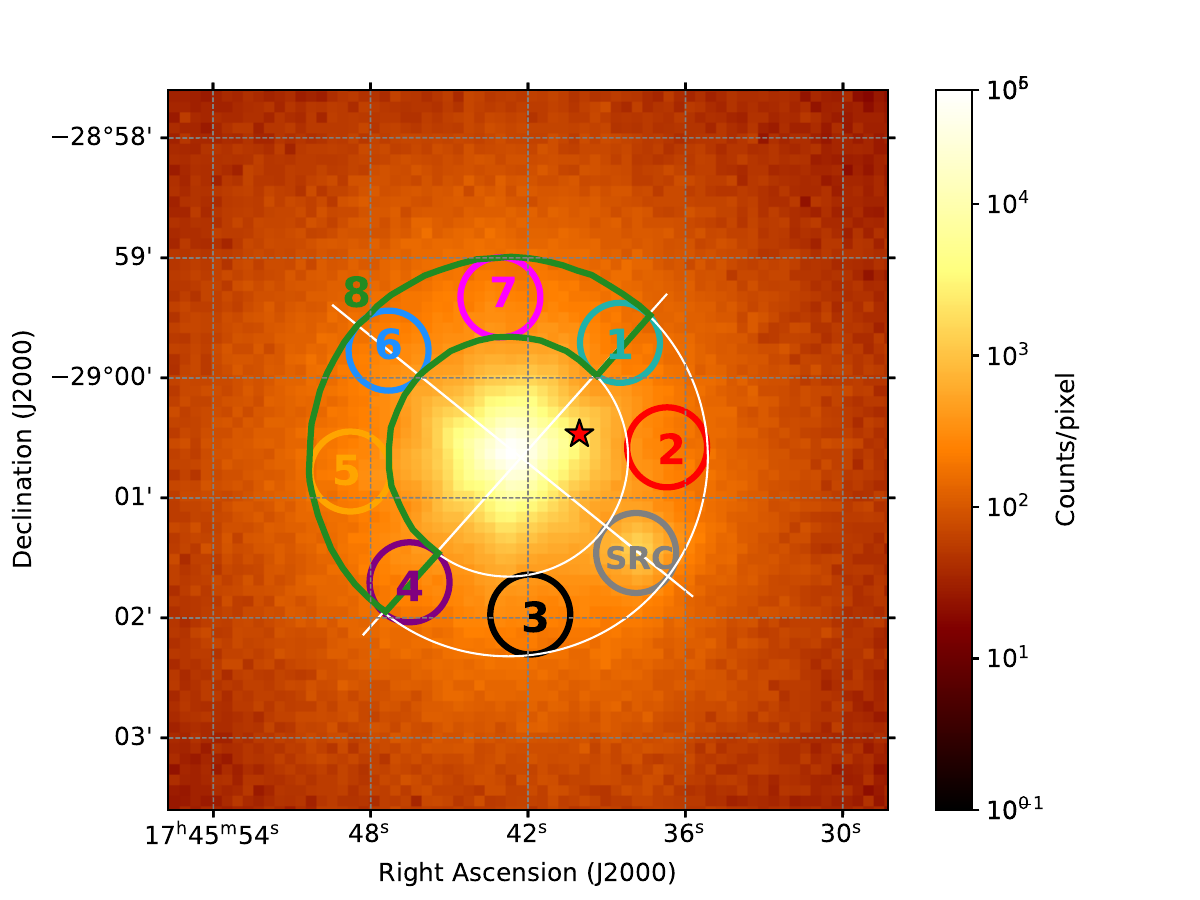}
\caption{IXPE image of \maxij (bright source in the center of the FOV) and \axj (SRC) in gray. The red star marks the position of Sgr~A$ ^{*} $. The selected circular background regions are numbered 1 to 7. The dark green semicircular annulus represents the 8th region, chosen to be diametrically opposite of the source region, as highlighted by the white guidelines. The two guidelines intersect the center of \maxij, defined as the pixel with the highest count rate.
\label{fig:regions}}
\end{figure*}

\section{Observation and Data reduction}
\label{sec:methods}

\axj is a transient low-mass X-ray binary system that was serendipitously observed in the \ixpe field of view during a DDT observation of \maxij \citep{Marra2025}. The observation started 2025 April 5 at 07:36:34 UTC and ended on April 8 at 02:09:33 UTC, for a total livetime of $ \approx $ 150~ks, with \textsc{obsid}: 04250301. We downloaded Level 1 and Level 2 data from the public HEASARC data archive.\footnote[1]{\url{https://heasarc.gsfc.nasa.gov/docs/ixpe/archive/}} We adjusted the timing of the event arrival relative to the solar system's barycenter applying the {\tt barycorr} tool from the \textsc{ftools}  package and performed background rejection according to \cite{DiMarco+23}, as \axj is a relatively low-brightness source. 

We chose a circular region centered on \axj and of radius 20\arcsec, to limit the extraction to the region in which the source of our interest is significantly brighter than the contamination from \maxij{}. The source count rate is $\approx 0.22$ count~s$^{-1}$~arcmin$^{-2}$, per Detector Unit (DU). The proximity of \maxij~ and Sgr~A$ ^{*} $, at a mere $\lesssim$1\farcm5 from \axj, makes it challenging to disentangle the source emission from the different background contributions. We could not extract an annular background region centered around \axj due to the high contamination by \maxij, which was observed with a much larger count rate of $\approx 1.14 \, \mathrm{count\,s^{-1}\, arcmin^{-2}}$ in the selected region. Therefore, we examine several possibilities, as depicted in Figure~\ref{fig:regions}. There we show the \axj source region marked ``SRC'' and our background regions marked ``1''--``8''. The regions are placed as follows: we identify the pixel with the highest count rate of \maxij and establish it as the center of the white ring with inner and outer radii of 60\arcsec\ and 100\arcsec, respectively. For the source region, we select a circle with a radius of 20\arcsec\ centered on \axj's position, within the white ring. We place seven evenly distributed circular regions of the same radius along the ring, noted as ``1''--``7''. The last background region of choice is a semicircular annulus, marked ``8'', positioned diametrically opposite the source region. We explore the effect of using each of these regions independently as a distinct background in order to assess the robustness of our conclusions.

The \textsc{SAOImageDS9} tool and the \texttt{xpselect} feature of the \ixpeobssim software (v. 31.0.1; \citealt{Baldini+22}) were consecutively used to respectively define the source and background regions and to select the studied region. The \texttt{xpbin} tool within the \ixpeobssim package was then used to generate polarization cubes (through the \texttt{pcube} algorithm) and to create the Stokes spectra $I$, $Q$, and $U$. Lastly, we used the \textsc{ftools} \texttt{ftgrouppha} task to rebin the background-subtracted Stokes spectra to require 3 counts per bin for Stokes $I$ and 5 counts per bin for $Q$ and $U$. We applied unweighted (\texttt{pcube}) and weighted (Stokes $I$, $Q$, $U$) analysis using the v. 20240701\_v013 response matrices, which is the latest version provided with \ixpeobssim\ v 31.0.1. A model-dependent analysis was performed on the binned spectra using the \textsc{heasoft} package \textsc{xspec} software \citep[version 12.15;][]{Arnaud+96}. For phase-folded analysis, we extracted light curves using the \textsc{ftools} task \texttt{xselect}, then \texttt{lcmath} to subtract background from light curves and sum the three DUs. In order to fold the light curves and obtain the orbital profile, the \texttt{efold} tool was used. Next, \texttt{xselect} was used to filter the orbital phases that correspond to eclipse and dip.  Finally, \texttt{xpbin} tool from \ixpeobssim  was applied to extract pcubes from the phase-filtered data.

\section{Results}\label{sec:results}

\subsection{Source and background polarization} \label{sec:pcube}

\begin{figure}
\centering 
\includegraphics[width=0.5\textwidth]{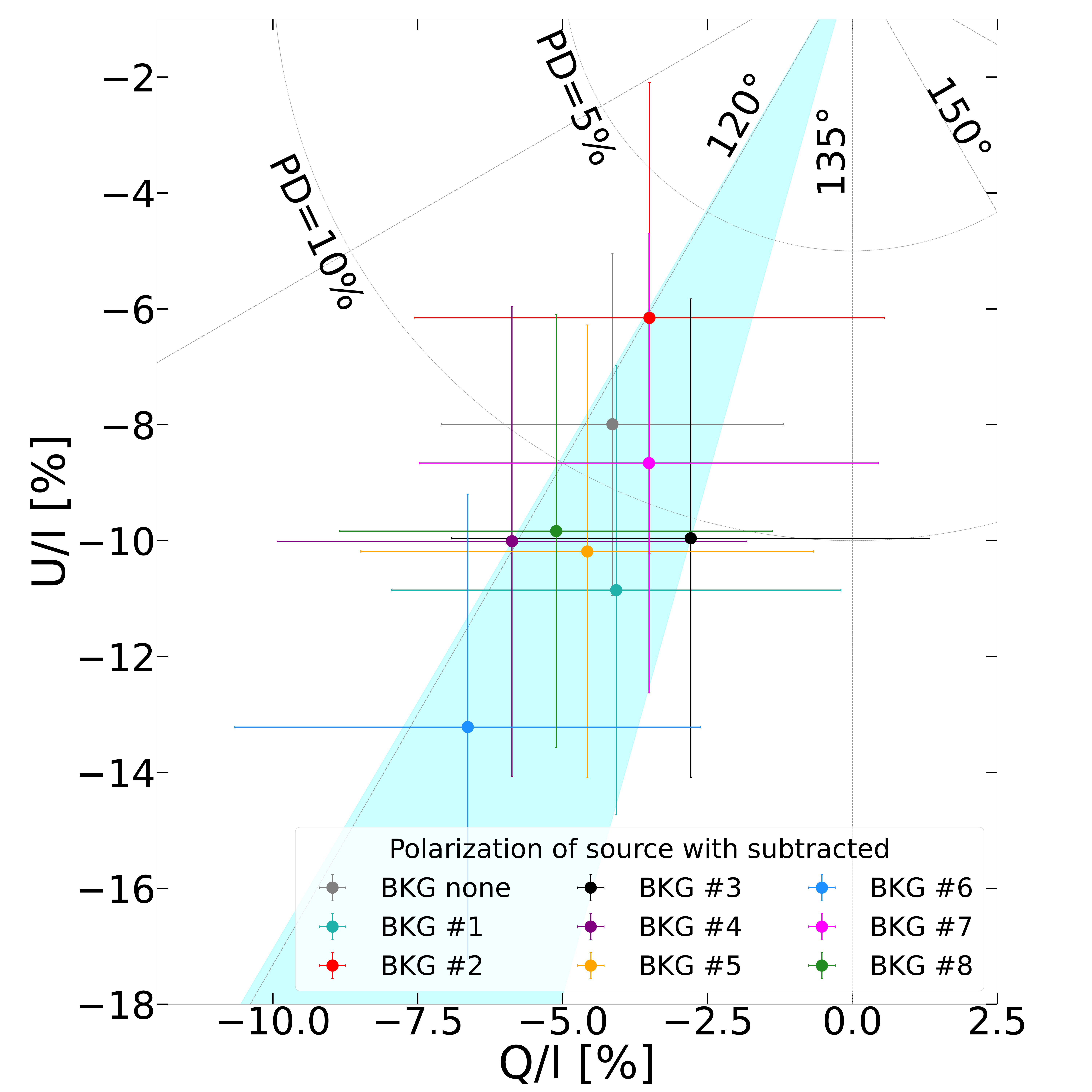}
\caption{Normalized Stokes parameters $U/I$ vs. $Q/I$ of the source with different background subtractions. The gray point labeled as ``BKG none'' refers to the polarization of the source region without background subtraction. The values of PD and PA from the combined data of the three DUs, for each background can be read using the gray grid. The points (plotted with 1 $ \sigma $ errors) are clustered within an angular wedge of size $ \approx 10\degr$, regardless of the background region subtracted. The clustering is represented by the cyan wedge. The gray grid represents the PD (in \%) and PA (in degrees).
\label{fig:pcube_src}}
\end{figure}

To ascertain the model-independent polarization, we applied the \texttt{pcube} algorithm of \ixpeobssim, obtaining the energy-averaged polarization degree (PD) and polarization angle (PA) shown in Figure~\ref{fig:pcube_src}. First, we assessed the source itself (no background subtracted) and found the PD = 8.9\%$\pm $2.9\% and PA = 121\degr$\pm$9\degr. The tentative PD (if considering two decimal places) lays above the minimum detectable polarization (MDP) value at the 99\% confidence level \citep{Weisskopf2010, Strohmayer2013} of 8.9\%, hinting that the source could be polarized. 
However, this signal is likely diluted by the strong, unpolarized emission from \maxij, for which a $3\sigma$ upper limit on the PD of 1.3\% was derived by \cite{Marra2025}. To investigate this further, we therefore examined the polarization properties of the source after subtracting different background regions individually, as illustrated in Figure~\ref{fig:regions} and explained in Section~\ref{sec:methods}. We report the measured PD and PA of the source with respective MDP values (where applicable) in the left panel of Table~\ref{tab:pcube}.

However, to properly interpret these background-subtracted measurements, we must first verify our assumption that the background regions are unpolarized. Given the crowded nature of the Galactic Center, we cannot exclude the presence of unresolved polarized sources contained in the selected background regions.

After calculating the \texttt{pcube} of the background regions (right panel of Table~\ref{tab:pcube}), it is clear that only one of the background regions (``2'') appears significantly polarized, as its PD~=~15.8\%~$\pm$~5.2\% is slightly above MDP = 15.7\%. This region is also the closest one to Sgr A$^{*}$, so it could harbor some contamination. To further examine our polarization measurements independently of background subtraction effects, we tested another diagnostic: the consistency of the PA values across different background region subtractions.

\begin{table*}

\begin{subtable}{.49\linewidth}
\begin{tabular}[t]{cccc}
\hline
\hline
Subtracted & PD [\%] & PA [deg] & MDP [\%] \\ 
BKG & & & \\
\hline

none & 8.9 $ \pm $ 2.9 & 121 $ \pm $ 9 & 8.9 \\

1 & 11.5 $ \pm $ 3.9 & unconstrained ($\sim125)$ & \\

2 & 7.0 $ \pm $ 4.1 & unconstrained ($\sim120)$ & \\

3 & 10.3 $ \pm $ 4.1 & unconstrained ($\sim127)$ & \\

4 & 11.5 $ \pm $ 4.0 & unconstrained ($\sim120)$ & \\

5 & 11.2 $ \pm $ 3.9 & unconstrained ($\sim123)$ & \\

6 & 14.7 $ \pm $ 4.0 & 122 $ \pm $ 7.8 & \\

7 & 9.4 $ \pm $ 4.0 & unconstrained ($\sim124)$ & \\

8 & 11.1 $ \pm $ 3.7 & 121 $ \pm $ 10 & \\

\hline
\end{tabular}
\end{subtable}
\begin{subtable}{.49\linewidth}
\begin{tabular}[t]{cccc}
\hline
\hline
Studied & PD [\%] & PA [deg] & MDP [\%] \\  
BKG & & & \\
\hline
none & - & - & - \\

1 & 6.0 $ \pm $ 5.5 & unconstrained ($\sim69)$ & 16.7 \\

2 & 15.8 $ \pm $ 5.2 & 123 $ \pm $ 9 & 15.7 \\
             
3 & 8.7 $ \pm $ 4.9 & unconstrained ($\sim95)$ & 14.9 \\

4 & 2.1 $ \pm $ 4.9 & unconstrained ($\sim167)$ & 14.9 \\

5 & 2.5 $ \pm $ 5.1 & unconstrained ($\sim81)$ & 15.6 \\

6 & 12.7 $ \pm $ 5.5 & unconstrained ($\sim33)$ & 16.6 \\

7 & 8.5 $ \pm $ 5.1 & unconstrained ($\sim110)$ & 15.5 \\

8 & 0.7 $ \pm $ 1.8 & unconstrained ($\sim124)$ & 5.5 \\

\hline
\end{tabular}
\end{subtable}
\caption{Left: Polarization degree and angle of \axj after subtracting various background regions shown in Figure~\ref{fig:regions} with ``none'' corresponding to the polarization of the source without background subtraction. MDP value is listed for the source before background subtraction. For values where the PD value is lower than 3$\times$ its uncertainty, we consider the PA unconstrained. Right: Polarization degree, angle, and MDP values of respective background regions shown in Figure~\ref{fig:regions}. The reported errors correspond to 1~$\sigma$. For unconstrained PA cases the most probable value is indicated.}
\label{tab:pcube}
\end{table*}

\begin{figure}
\centering 
\includegraphics[width=0.5\textwidth]{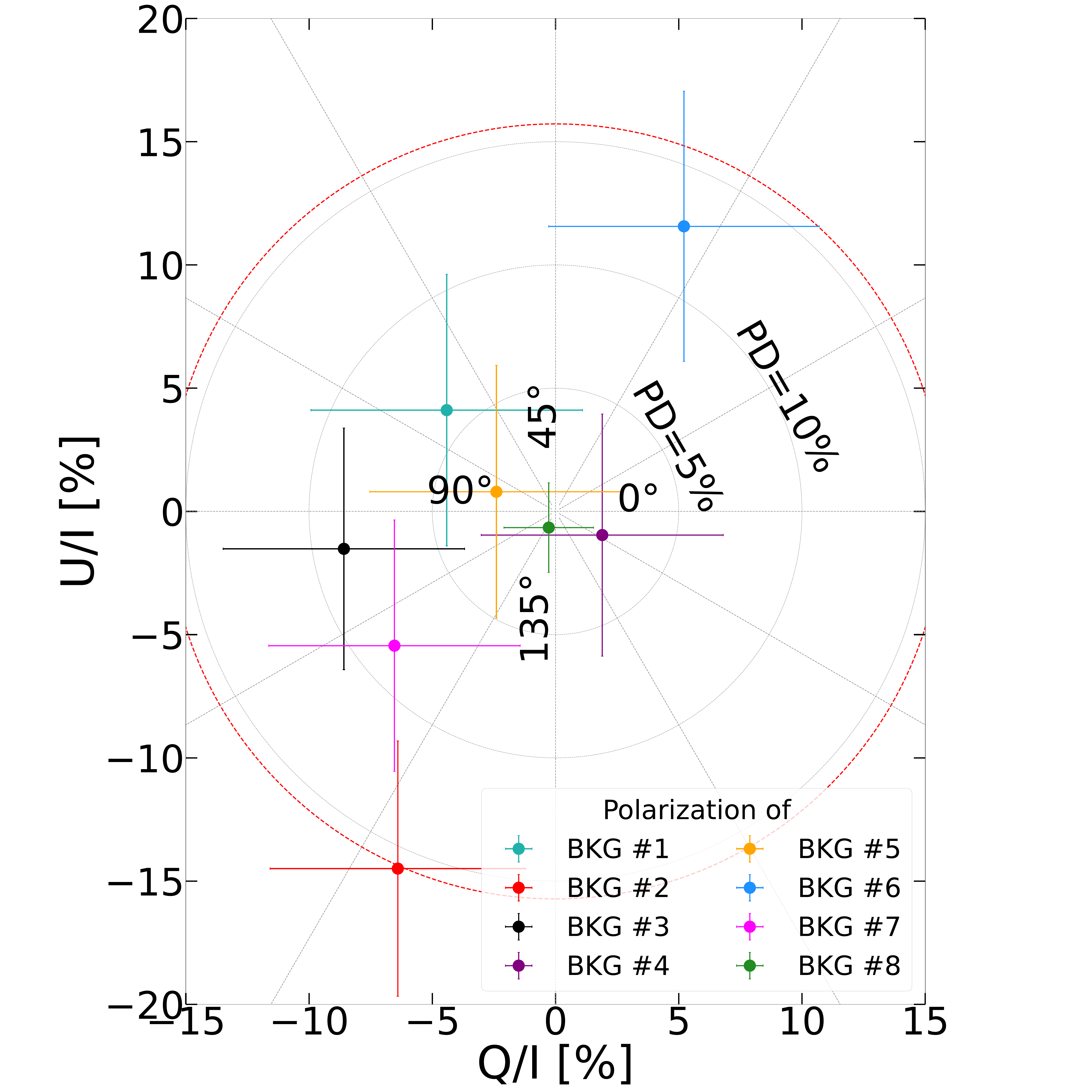}
\caption{Normalized Stokes parameters $U/I$ vs. $Q/I$ of different background regions.
The values of PD and PA of the sum of the three DUs, for each background, can be read using the gray grid. The only region with a PD above MDP is region ``2'' (in red), which is probably polarized. All other regions are well below the MDP and have scattered PA. The gray grid represents the PD (in \%) and PA (in degrees). The red circle represents the MDP value for background region ``2''.
\label{fig:pcube_bkg}}
\end{figure}

Another clue hinting towards the source being polarized comes from the consistency of PA values across different background subtractions. For a truly polarized source, the measured PA should remain consistent regardless of which background region is subtracted. If the PA values vary significantly between different background subtractions, it suggests the signal may be dominated by noise rather than true polarization. In Figure~\ref{fig:pcube_src}, we show the normalized Stokes parameters $U/I$ vs. $Q/I$ for the source after different background region subtractions. Using the secondary (gray) grid, we can also read the values of PD and PA. It is a robust sign that the PA stays consistent irrespective of the background subtracted. The dispersion of the measured PA values stays within the wedge of $ \Delta$PA$\sim 10\degr$. The background regions, on the contrary, show a rather ``non-concentrated'' distribution of PA values (Figure~\ref{fig:pcube_bkg}).

\begin{figure*}
\centering \includegraphics[width=0.7\textwidth]{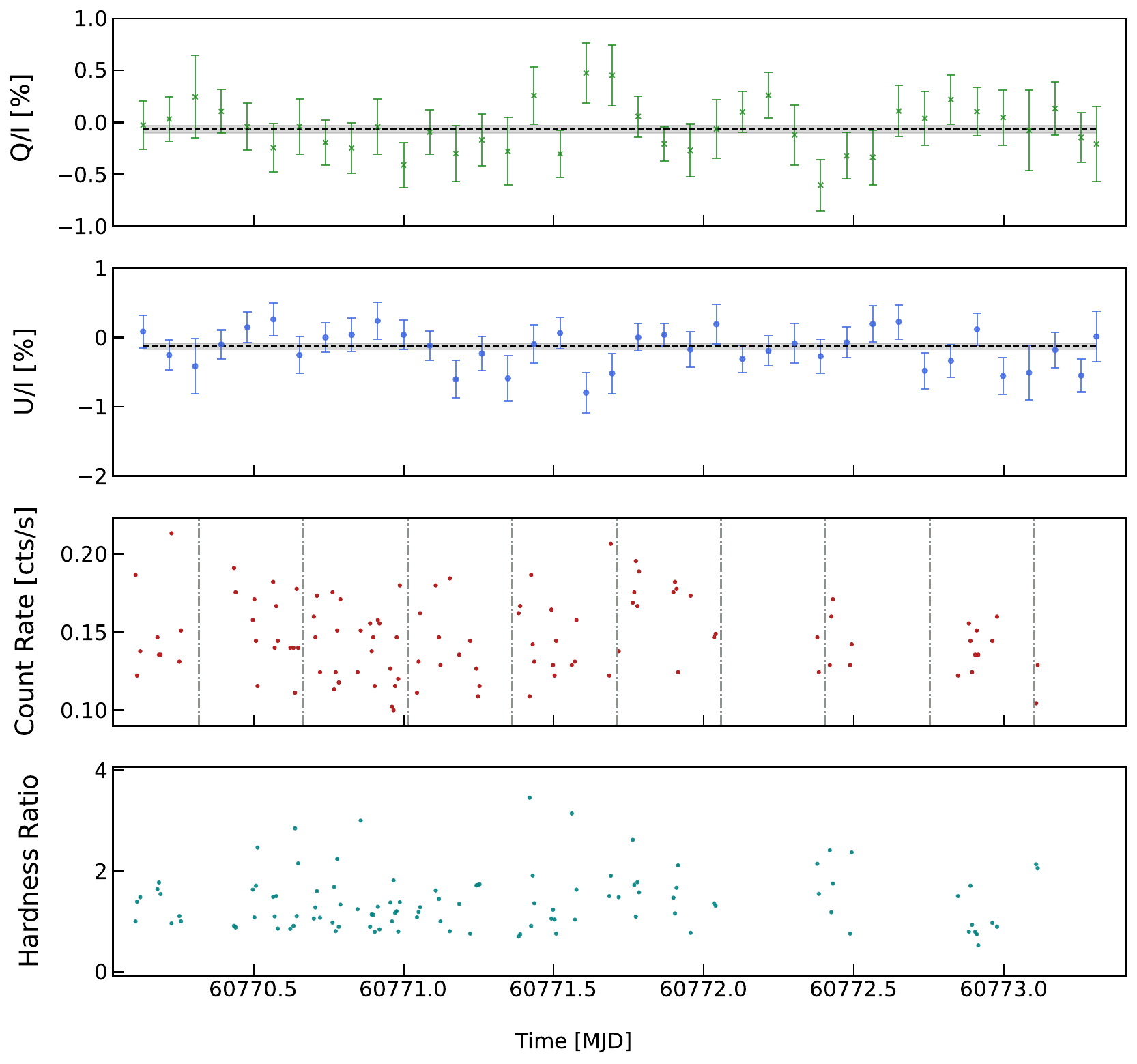}
\caption{First and second sub-panels (respectively): normalized Stokes parameters $Q/I$ and $U/I$ as a function of time. The data are binned in intervals of 7.5 ks and are energy-averaged over the 2--8~keV band. The dark gray dashed line represents the constant fit, its 1$ \sigma $ error is represented by the shaded area. Third sub-panel from top: time evolution of count rate in the 2--8~keV band. Gray lines represent expected eclipses. Fourth sub-panel: time evolution of hardness ratio defined as a ratio between 4--8~keV and 2--4~keV count rates. The count rate and hardness ratio show data binned in intervals of 300~s. Each data point in all subplots represents the sum of the 3 DUs.
\label{fig:QN_UN_LC_HR}}
\end{figure*}

Considering the complexity of the Galactic Center and the fact that it is highly populated with sources, it is a difficult task to choose the ``right'' background region. The logical step would be to consider a region that is diametrically opposite to the source region (for symmetry reasons), and that contains the least number of \axj source counts possible (to limit self-subtraction). This leaves us with only two choices: regions ``6'' and ``8''. Since the latter spans a vast area, it poses a higher chance of including an uncontrolled effect, such as containing an unresolved source (i.e., a source that we cannot distinguish but is polarized). Part of region ``8'' is also in the vicinity of Sgr A$ ^{*} $, and we cannot exclude possible contamination. Background region ``6'' however, satisfies both conditions of being far from Sgr A$^{*} $ and being in a reasonable position with respect to both \maxij and \axj. 
Moreover, this region is the most logical choice to subtract possible systematic effects which impact the spatial response to polarization of photoelectric polarimeters. These are present when the extraction region is not uniformly illuminated and manifest themselves as a polarization along the image gradient \citep{Soffitta2013, Bucciantini2023}. In our case, such systematic effects would add a radial polarization directed from the central bright spot (\maxij), with a higher PD when the gradient is steeper. Therefore, in region ``6'', this effect would be the same in direction and amplitude as in the source region and then any spurious contribution to polarization which may be present would be removed in the background subtraction process. 
For these reasons, we will proceed with our study of \axj considering only the background region ``6''\footnote[2]{While our choice of background ``6'' is motivated by reducing systematic effects, choosing the alternative background region ``8'' also provides a significant detection of polarization.} 
and its subtraction from the source in order to study the polarization properties of \axj.

\subsection{Time-dependent analysis}

In this subsection, we explore the temporal evolution of both the polarimetric and spectral properties of the source, summarized in Figure \ref{fig:QN_UN_LC_HR}. We plotted the normalized Stokes parameters $ Q/I $ and $ U/I $ in the first and second sub-panels, respectively. The data shown are binned in 7.5-ks intervals over the entire observation duration. The third and fourth sub-panels from top show the count rate with expected eclipses, and the hardness ratio, respectively, binned in 300 s intervals each. We analyzed the first two datasets (Stokes parameters $ Q/I $ and $ U/I $) to test for possible time variability of the polarimetric properties of the source. The best fit we obtained was using a constant model (gray dashed lines in the figure). $ Q/I $ and $ U/I $ are two independent variables following a normal distribution \citep{Kislat+15}. This allows us to sum their $ \chi^2 $ and degrees of freedom in order to test the null hypothesis. We obtained $ \chi^2 = 73 / 74$ d.o.f and null hypothesis probability of 51\% for the constant model. We conclude that the polarimetric properties stayed constant over the duration of the observation, although the high statistical uncertainties of our data limit our analysis.

\begin{figure} 
\hspace*{-0.5cm}
\includegraphics[width=0.48\textwidth]{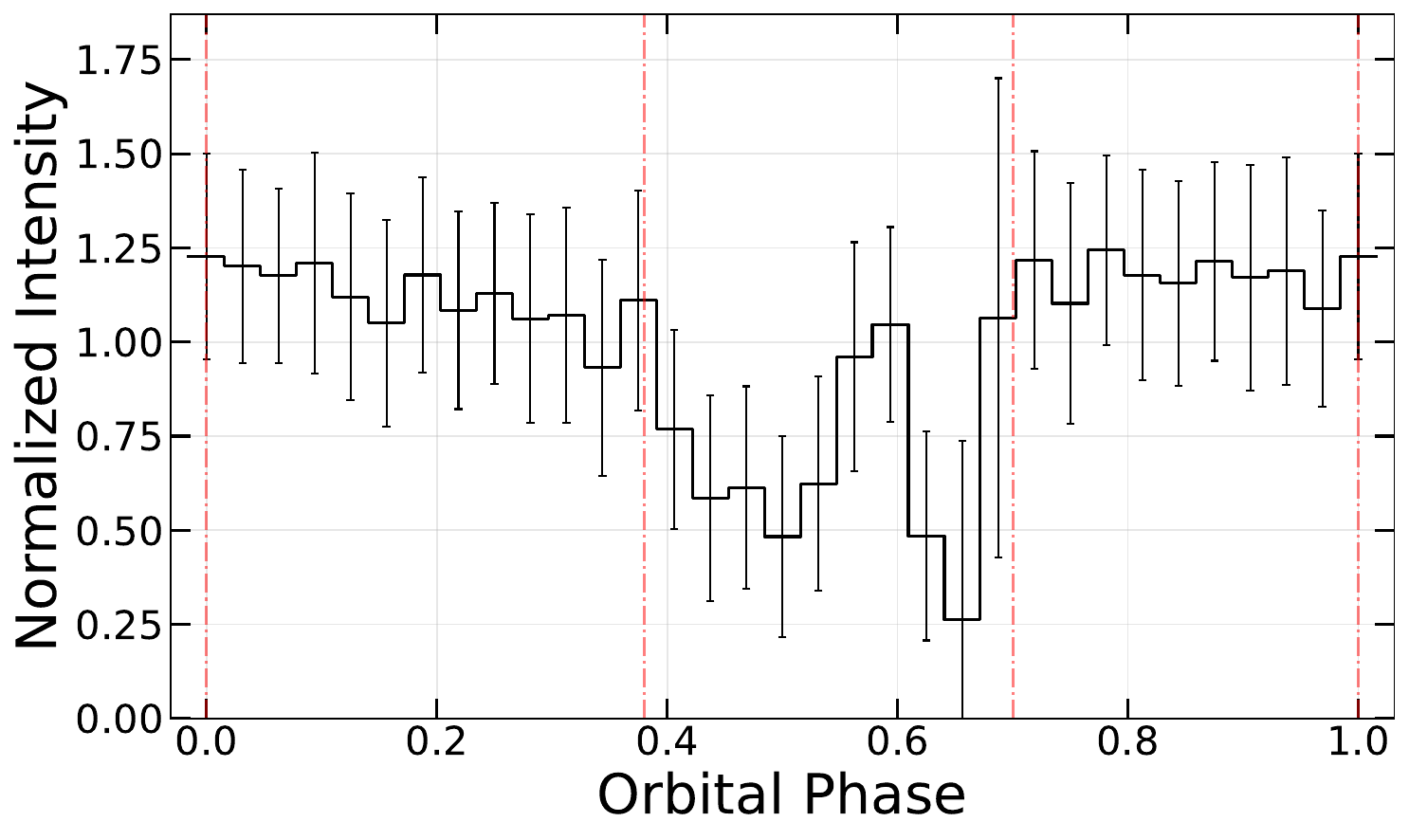}
\caption{Phase versus normalized intensity. The plot is divided into three parts separated by red dash-dotted lines: (i) before the dip + eclipse (phase 0--0.38), (ii) dip + eclipse (phase 0.38--0.70), (iii) after the dip + eclipse (phase 0.70--1.).
\label{fig:phase}}
\end{figure}

\subsection{Phase-resolved polarimetry}

\begin{figure} 
\hspace*{-0.5cm}
\includegraphics[width=0.48\textwidth]{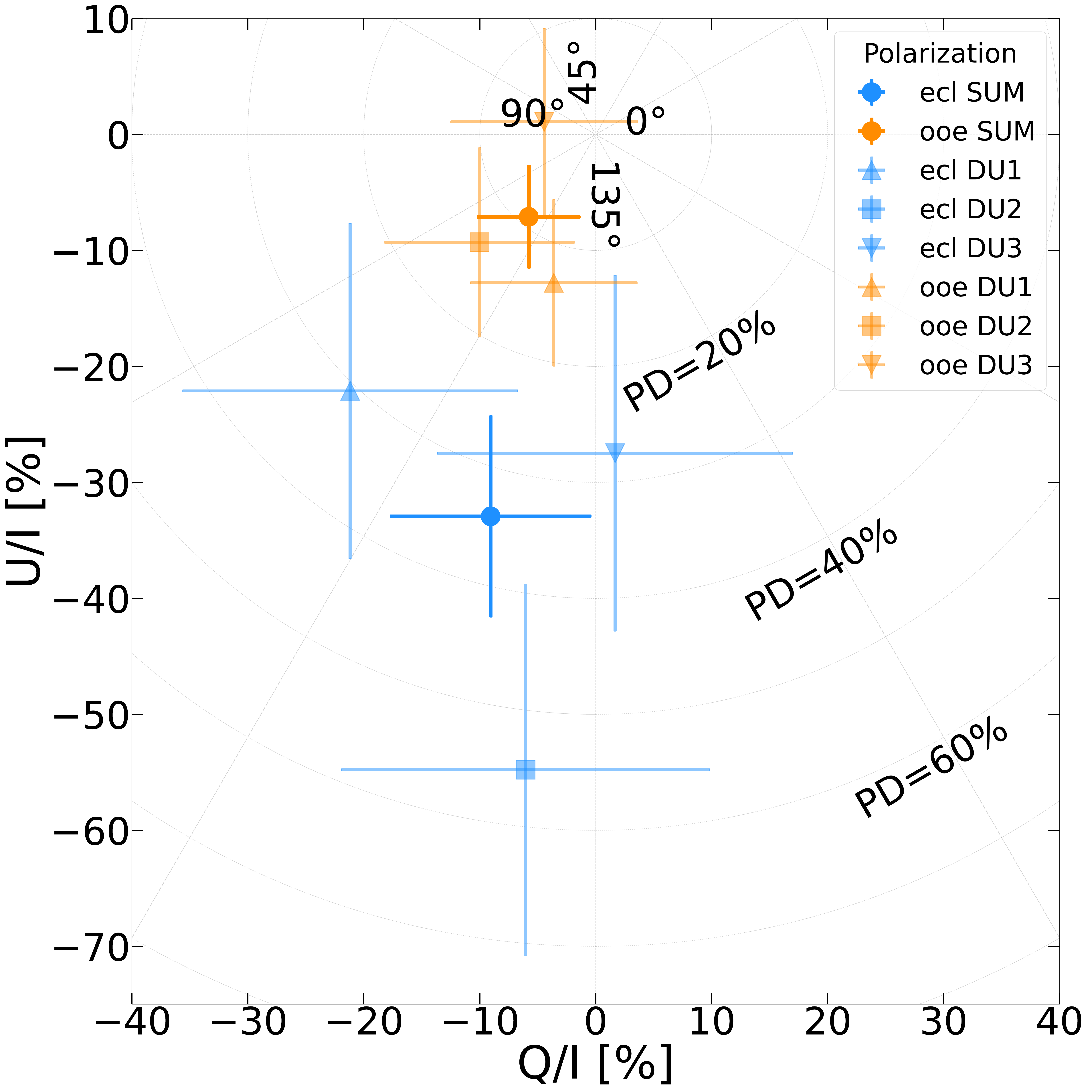}
\caption{Normalized Stokes parameters $U/I$ vs. $Q/I$ of the source for different phase bins. The blue points represent the dip + eclipse (ecl), corresponding to phase 0.38 -- 0.70 in Figure~\ref{fig:phase}, while orange points represent the out-of-eclipse (ooe) phase, corresponding to combined regions 0 -- 0.38 and 0.70 -- 1. We plot a value integrated over the three DUs (labeled as SUM), as well as each DU separately (DU1, 2, and 3). The gray grid represents the PD (in \%) and PA (in degrees).
\label{fig:pcube_phase}}
\end{figure}

In the next step, we examine the periodic behavior of the source, focusing on the variations of the observed polarimetric properties. Using the orbital solution from \cite{Ponti2018}, we folded the background-subtracted, summed over the three DUs, lightcurves at an arbitrary epoch into 32 bins. The dip and eclipse in the orbital profile were visually identified (Figure~\ref{fig:phase}). We then extracted the source and background event files for each DU in two phase intervals, which are 0.38--0.70 (dip + eclipse) and the complementary phases. Polarization cubes were constructed from each event file. In the phase region of the dip and eclipse, we measured PD of 34.2\% $\pm$ 8.7\%. The corresponding PA was 127\degr $\pm$ 7\degr, which is consistent with the time-averaged result (see Figure~\ref{fig:pcube_phase}). In contrast, in the phase interval outside the dip + eclipse, we found PD of 9.2 $\pm$ 4.5\%.

We observe a remarkably high PD during the phase with a dip and eclipse. While higher polarization during dip and/or eclipse is expected for high-inclination systems, the measured value of 34.2\% is surprisingly large. This suggests that during the eclipse, the companion star obscures the central source, absorbing much of the unpolarized emission, while polarization is likely produced by scattering. Similarly, we expect that dips are related to partial obscuration by the outer regions of the disk. The scattering process is likely responsible for high polarization observed in \axj. The reflecting medium is possibly either accreting disk winds or outflow similar to that observed in Cygnus X-3 \citep{Veledina+24, Veledina+24_Soft}.

\begin{figure} 
\hspace*{-0.5cm} 
\centering
\includegraphics[width=0.85\linewidth]{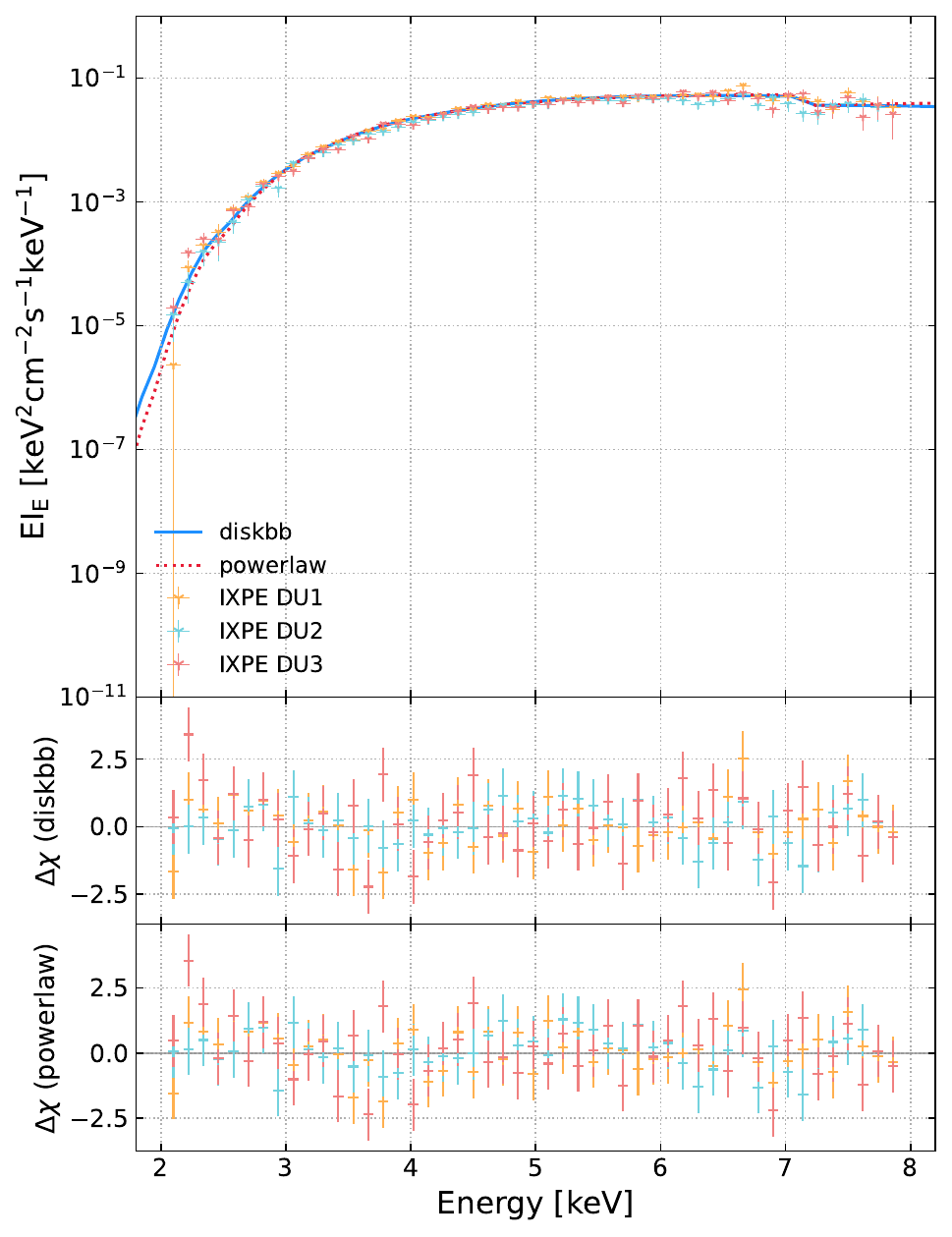}
\caption{X-ray spectra of \axj observed with \ixpe per each DU. Upper panel: unfolded $EF_E$ spectra of Stokes $I$, comparing goodness of fit for a model with a disk blackbody (solid line) and a power law (dotted line). Lower panels: residuals of the fit $\Delta \chi$  with the disk blackbody and power-law models, respectively.
\label{fig:fit}}
\end{figure}

\begin{table*}
\begin{center}
\begin{tabular}{p{2cm}p{3cm}p{4.5cm}p{3cm}p{3cm}}
\hline
\hline
Component & Parameter [unit] & Description & Value Fit 1 & Value Fit 2 \\  
\hline
{\tt constant} & C$_{\mathrm{DU1}}$ & Normalization & \textbf{1.00} & \textbf{1.00} \\
               & C$_{\mathrm{DU2}}$ & Normalization & $ 0.86 \substack{+0.04 \\ -0.03} $ & $ 0.86 \substack{+0.04 \\ -0.03} $ \\
               & C$_{\mathrm{DU3}}$ & Normalization & $ 0.91 \substack{+0.04 \\ -0.03} $ & $ 0.91 \substack{+0.04 \\ -0.03} $ \\
{\tt TBabs} & $N_\textrm{H}$ [$10^{22}$\,cm$^{-2}$] & Hydrogen column density & $ 22.3 \pm 1.2 $ & $ 26.3\substack{+1.8\\ -1.7}$ \\
{\tt polconst} & $ A $ [\%] & Polarization degree & $ 15.4 \pm 5.6 $ & $ 15.3 \pm 5.6 $ \\
             & psi [deg]    & Polarization angle                 & $ 120 \pm 11 $ & $ 120 \pm 11 $ \\
{\tt diskbb} & $k T_{\rm bb}$ [keV] & Inner disk radius temperature & $ 1.75 \substack{+0.15 \\ -0.13} $ & $ - $ \\
             & norm     & Normalization                 & $ 1.51\substack{+0.73 \\ -0.49}$ & $ - $ \\          
{\tt powerlaw} & $\Gamma$ & Photon index & $ - $ & $ 3.0 \substack{+0.3 \\ -0.2} $ \\
               & norm     & Normalization & $ - $ & $ 0.57\substack{+0.36 \\ -0.21} $ \\
              & $\chi^2$/d.o.f. &  & $ 304 / 314 $ & $ 316 / 314 $ \\ 
\hline
\end{tabular}
\caption{Best-fit parameters of the {\it IXPE} spectral fitting using the model described in Section~\ref{subsec:spectra}. Uncertainties are reported at 90\% confidence level. The parameters in bold were frozen during the fit.}
\label{tab:spectralfit}
\end{center}
\end{table*}

\subsection{Spectral properties}
\label{subsec:spectra}

\begin{figure} 
\hspace*{-0.5cm}
\centering
\includegraphics[width=0.90\linewidth]{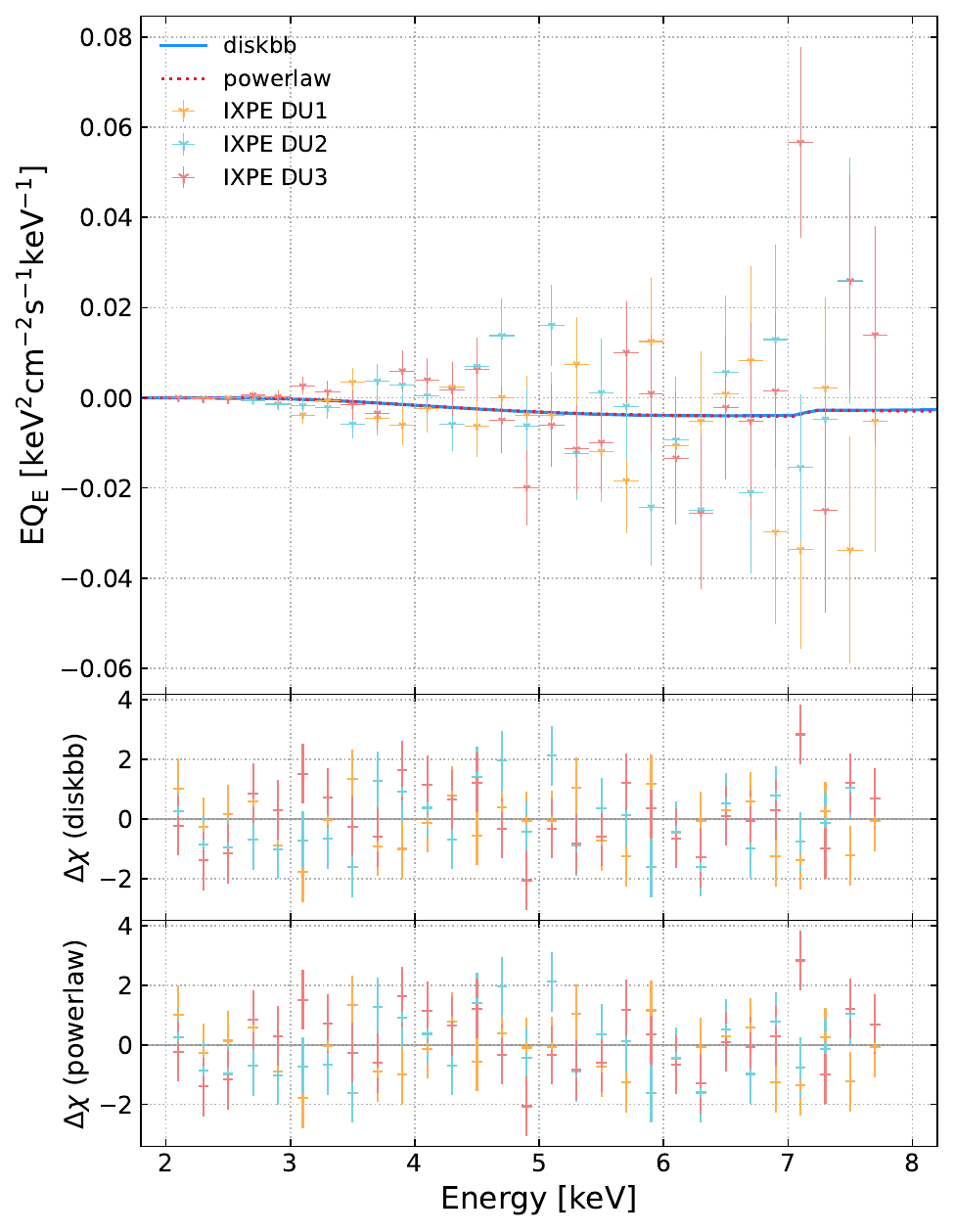}
\caption{Same as Figure~\ref{fig:fit}, but for the Stokes $Q$.}
\label{fig:fitQ}
\end{figure}

\begin{figure} 
\hspace*{-0.5cm}
\centering
\includegraphics[width=0.90\linewidth]{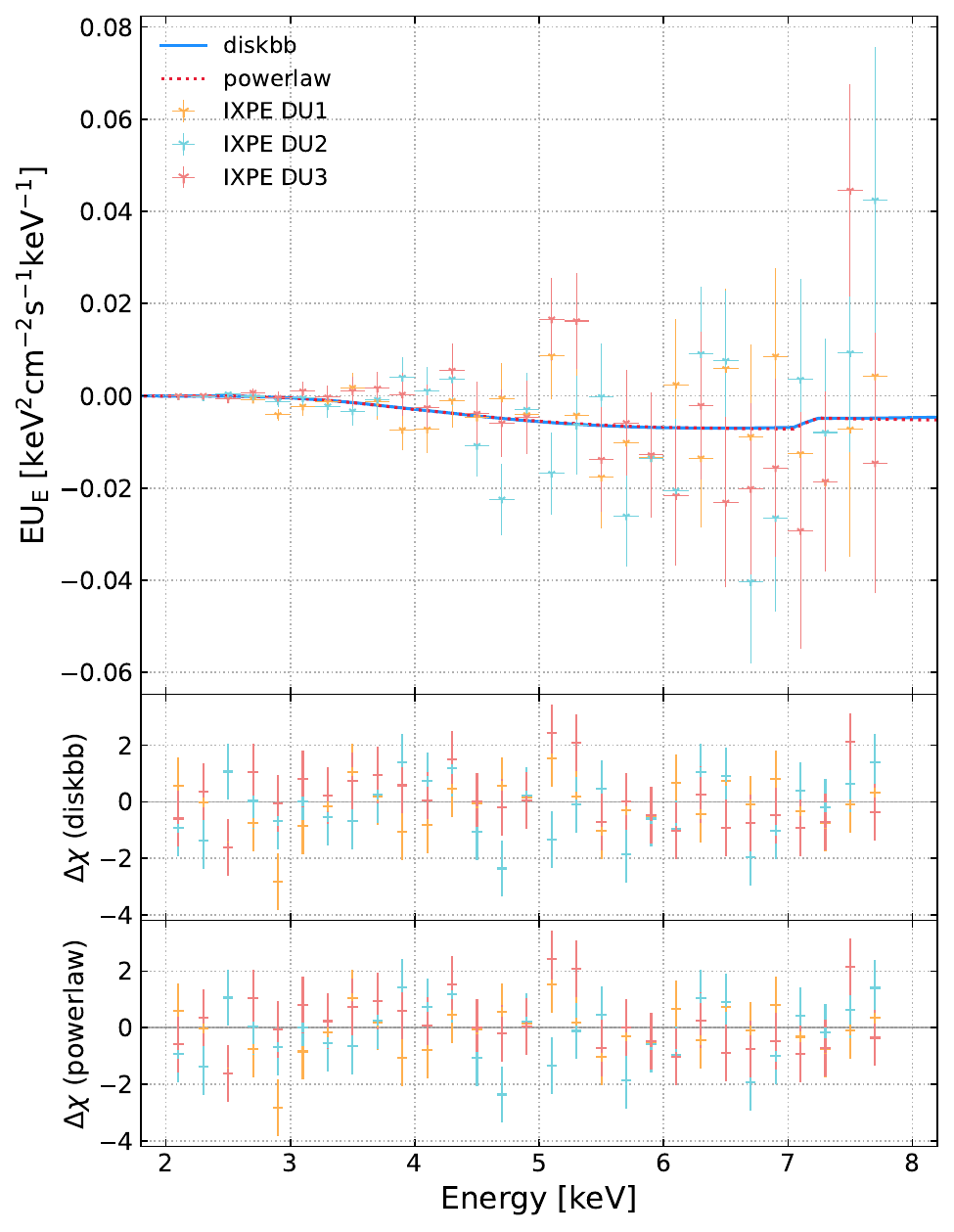}
\caption{Same as Figure~\ref{fig:fit}, but for the Stokes $U$.}
\label{fig:fitU}
\end{figure}

In our study of the source, we are also interested in disentangling the processes contributing to its X-ray emission. For this, we opt for comparing two simple models against each other: a disk blackbody versus a power law. The full model is:
\texttt{constant $ \times $ TBabs $ \times $ polconst $ \times $ emission}, 
where \texttt{constant} accounts for different normalization between the three DUs of the telescope, \texttt{TBabs} is the Galactic absorption, \texttt{polconst} is a polarization model assuming constant properties in the 2--8 keV range, and \texttt{emission} is either \texttt{diskbb} (black-body accretion disk model) or \texttt{powerlaw} (photon powerlaw). The best-fit parameters are reported in Table~\ref{tab:spectralfit} and the fits with the data residuals are shown in Figures~\ref{fig:fit}--\ref{fig:fitU}, where we plot the contribution of each single DU and compare the two models. The two fits with \texttt{diskbb} or \texttt{powerlaw} components give $\chi^{2}$/d.o.f. = 304/314  (null hypothesis probability = $ 0.64 $) and $\chi^{2}$/d.o.f. = 316/314 (null hypothesis probability = $ 0.46 $), respectively. While in both cases we obtain statistically good results, we cannot reject either model purely on the basis of the quality of fit. This can be seen from the top panel of Figure~\ref{fig:fit} where the two models overlap in the 3--7~keV energy band, and the corresponding residuals are plotted in the middle and bottom panels. While for the Stokes parameter $I$ there is only a minor difference around  2--2.5~keV and 7.5--8~keV, for both $Q$ and $U$ it is practically impossible to distinguish between the two tested models, as clear from Figures~\ref{fig:fitQ} and \ref{fig:fitU} for $Q$ and $U$, respectively. The fact that we do not manage to accurately distinguish between the two tested models is mainly due to only fitting the \ixpe data, which are observed in a relatively narrow energy range.

\begin{figure} 
\hspace*{-0.5cm}
\centering
\includegraphics[width=0.90\linewidth]{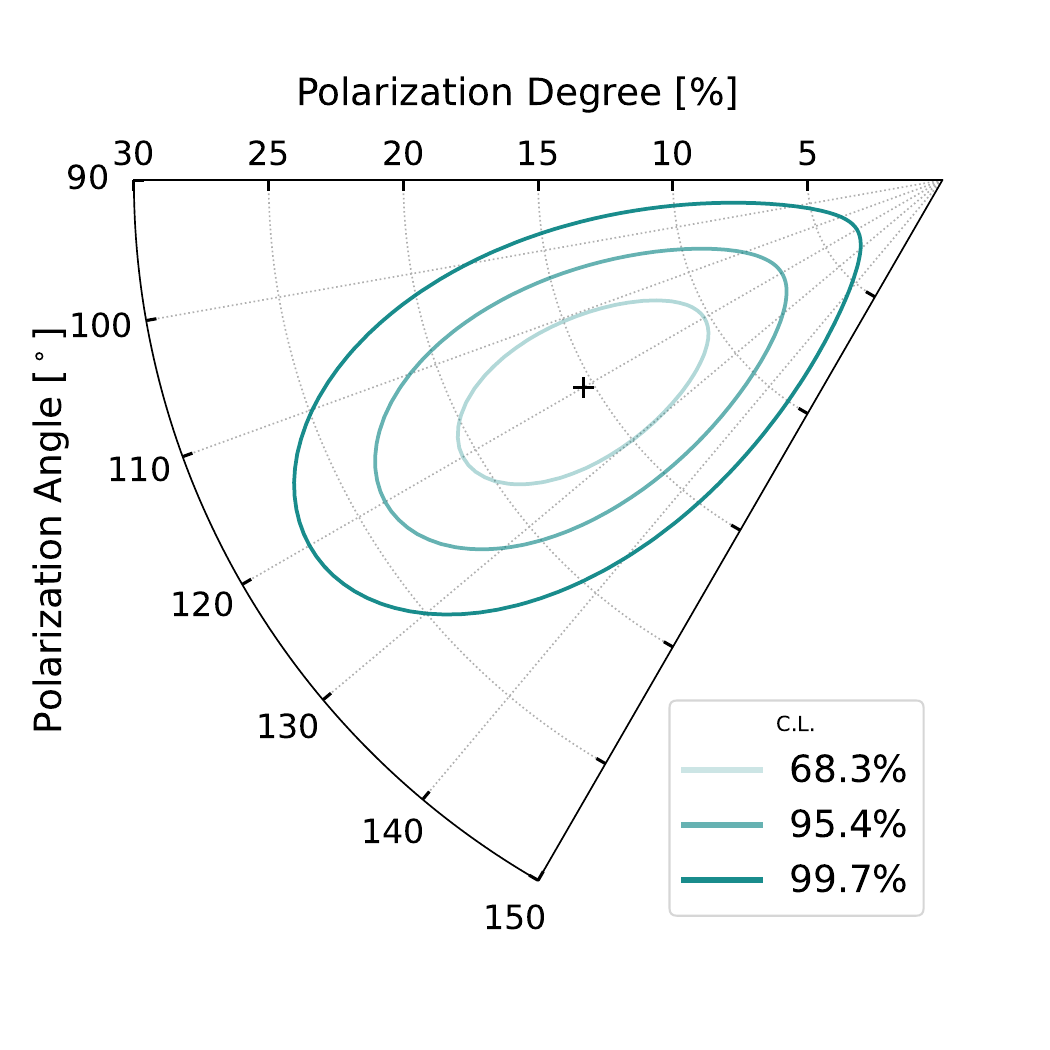}
\caption{Polar plot of PD and PA, assuming the best spectral fit for the \texttt{diskbb} model scenario. The black ``+'' sign represents the best-fit value for the given model, as reported in table~\ref{tab:spectralfit}. Results shown are considering the entire \ixpe\ energy band (2--8 keV). Colors are fainter with lower C.L.}
\label{fig:contour}
\end{figure}

Nevertheless, the spectral results we obtain are broadly consistent with the results obtained for the soft state by \cite{Ponti2015}.
Despite the statistical similarity of the two tested models, it is important to notice that the photon index of the \texttt{powerlaw} $ \Gamma = 3.0 $ is incompatible with the hard state \citep[e.g.,][]{Ponti2015}. The measured flux by \ixpe in 2--8 keV band is $ F_{\rm 2-8~keV} = 5.1 \times 10^{-11}\, \rm{erg\,cm^{-2}\,s^{-1}} $, with a hardness ratio $ F_{\rm 4-8~keV} / F_{\rm 2-4~keV} = 8.3\, \substack{+0.1 \\ -0.8}$, which corresponds to flux $ F_{\rm 3-10~keV} = 6.03 \times 10^{-11}\, \rm{erg\,cm^{-2}\,s^{-1}} $ with a hardness ratio $ F_{\rm 6-10~keV} / F_{\rm 3-6~keV} = 0.97\, \substack{+0.02 \\ -0.11}$. The latter value of the hardness ratio is consistent with the historical measurement of the source in the soft state \cite[Fig. 3 in][]{Ponti2018}, while the flux would correspond to the area (in the same figure) between the hard and soft spectral states. It is possible that the source was transitioning between the two states during the \ixpe\ observation. We see from (Fig. 2 in the same paper), that the transition between the two states can take extended periods of time. Figure \ref{fig:contour} presents the PD and PA contours from the model-dependent analysis for the \texttt{diskbb} scenario. We have created the contour plot using the \texttt{steppar} command in \texttt{xspec} with 60 steps for both PD and PA. We find the PD and PA values consistent with the \texttt{pcube} result within the error margin.

\section{Discussion}\label{sec:discussion}

The serendipitous detection of \axj during the \ixpe observation of \maxij represents a valuable opportunity to conduct the first X-ray polarimetric study of this neutron star LMXB system. Located within $\sim$1\farcm5 of the Galactic Center, \axj provides a unique laboratory for investigating polarization signatures in high-inclination accreting systems, despite the observational challenges posed by its location in this crowded and complex region. While \axj is not the first highly inclined NS-LMXB observed by \ixpe, it presents distinctive characteristics that make it particularly interesting for polarimetric studies. Previous \ixpe observations of the highly inclined ($i \sim 60$--$ 80\degr$, \citealt{DiazTrigo2012, Tomaru2020}) dipping source GX~13+1 revealed an average polarization degree of 2--5\%, rising to 7.5\% $\pm$ 1.7\% during dips \citep{Bobrikova2024a, Bobrikova2024b, DiMarco2025, DiMarco2025WMNS}. More recently, observations of the eclipsing source 2S~0921--630, with an inclination of $i \sim 82\degr$ \citep{Ashcraft2012}, measured an average PD of 8.8\% $\pm$ 1.4\% \citep{Tomaru2025}, representing the highest polarization measured for NS-LMXBs prior to our \axj observation. The significantly higher polarization we detect in \axj thus adds an important new data point to this emerging picture of polarization in high-inclination neutron star systems.

Several observational challenges significantly limit the depth of our analysis. The proximity of the much brighter \maxij ($\sim$1\farcm3 away) creates substantial difficulties in background subtraction, as conventional annular background regions cannot be utilized without severe contamination, thus requiring a completely different approach. The extremely crowded nature of the Galactic Center region further complicates the selection of appropriate background areas, forcing us to rely on carefully chosen circular regions that may not perfectly represent the local background conditions around \axj. The intrinsic faintness of \axj (count rate $\approx$ 0.22 count s$^{-1}$ arcmin$^{-2}$) adds to the aforementioned difficulties, resulting in relatively large statistical uncertainties that limit our ability to constrain the polarization properties with high precision.

The temporal analysis did not show any significant variations of the polarimetric properties throughout the 150~ks observation. \axj exhibits clear eclipses and dipping behavior in X-ray intensity tied to its orbital phase \citep{Ponti2015, Ponti2018}. Our phase-resolved analysis reveals an increase of polarization degree from $ \approx $ 9\% outside of eclipse to $ \approx $ 34\% during dip and eclipse phase. This increase in polarization degree points towards scattering processes as polarization-inducing mechanisms, likely originating in disk winds.

The low signal-to-noise ratio also impacts our spectral analysis, where we find that both disk blackbody and power-law models provide statistically acceptable fits ($\chi^2$/d.o.f. = 304/314 and 316/314, respectively) with comparable goodness-of-fit statistics. While we cannot confidently distinguish between disk blackbody and power-law emission models based on spectral fitting alone, the high photon index of our power-law fit and the broad consistency of our model parameters, flux and hardness ratio with those reported by \citep{Ponti2015} for soft state observations suggest that \axj~ was likely in a soft spectral state during our observation. Under this soft state interpretation, the observed polarization signal can be more confidently attributed to geometrical asymmetries in the thermal emission components, such as disk inclination effects, boundary layer structure, or scattering in moderately ionized winds, rather than non-thermal processes associated with hard-state Comptonization.

The degeneracy between thermal disk emission and non-thermal power-law components is particularly problematic for understanding the accretion state and the mechanisms responsible for the observed polarization. In thermal-dominated states, polarization might arise from scattering in the disk atmosphere \citep{Chandrasekhar+60,Sobolev+63,Loskutov+81,Taverna+21} or its wind \citep{Nitindala2025}, from geometrical asymmetries and relativistic effects in the boundary layer \citep{Bobrikova2025}, or reflection of the spreading layer emission from the accretion disk \citep{Lapidus1985}, while power-law dominated emission could indicate Comptonization processes that would produce different polarization signatures \citep{Poutanen1996}. Because the \ixpe band is restricted to the narrow 2--8 keV energy interval, this degeneracy cannot be broken, limiting our ability to fully utilize the diagnostic potential of our polarimetric measurements.

The detection of a polarization signal in \axj is particularly significant given its exceptional properties. \axj is well-suited for polarimetric studies thanks to its high inclination and history of wind detection — characteristics that may enhance polarization signatures, as evidenced by recent observations of wind-bearing sources such as the black hole binaries 4U~1630$-$47 \citep{Ratheesh2024} and IGR~J17091$-$3624 \citep{Ewing2025}, which exhibited larger polarization than previously expected. Both reflection and winds have been reported for this source \citep{Ponti2018}, and these processes can contribute substantially to the polarimetric signal through scattering mechanisms. The high measured PD represents the first such measurement for this source and adds to the growing catalog of polarimetric observations of NS-LMXBs, with important implications for understanding highly inclined accreting systems.

\section{Conclusions}\label{sec:conclusions}

We present the first X-ray polarimetric observation of the neutron star LMXB \axj using \ixpe, serendipitously obtained during a 150 ks observation of the nearby transient \maxij. Our main findings are:

\begin{itemize}
\item \axj shows evidence for X-ray polarization with PD~=~14.7\%~$\pm$~4.0\% and PA~=~122\degr~$\pm$~8\degr. The consistency of the PAs across different background subtraction choices bolsters confidence in this detection (Figures \ref{fig:pcube_src} and \ref{fig:pcube_bkg}, and Table \ref{tab:pcube}).

\item No significant temporal evolution of the polarimetric properties were observed throughout the 150~ks observation (Figure \ref{fig:QN_UN_LC_HR}).

\item The measured PD of 14.7\% $\pm$ 4.0\% is exceptionally high for NS-LMXBs, making \axj the most polarized neutron star binary observed to date — significantly exceeding the $\sim$4--5\% typically seen in Z-sources and reaching levels previously observed only at high energies (7--8 keV). This high polarization is consistent with the known presence of both scattering and wind signatures in this source \citep{Ponti2018}, which can contribute substantially to the polarimetric signal through scattering processes.

\item The phase-resolved analysis shows extraordinarily high PD in the dip+eclipse phase of 34.2\%~$\pm$~8.7\%, while the complementary phases are unpolarized. Such a high value of PD during the eclipse and dip phase strongly suggests that emission is polarized in the disk wind via scattering mechanisms.

\item Spectral analysis reveals that both disk blackbody and power-law models provide equally good fits (Figures \ref{fig:fit}--\ref{fig:fitU}) to the data, preventing definitive identification of the dominant emission mechanism and limiting physical interpretation of the polarization signal. However, the high photon index value of the power-law component ($ \Gamma = 3.0 $, Table \ref{tab:spectralfit}) is in favor of a soft spectral state of the source.
\end{itemize}

\begin{acknowledgments}
The Imaging X-ray Polarimetry Explorer (IXPE) is a joint US and Italian mission.  The US contribution is supported by the National Aeronautics and Space Administration (NASA) and led and managed by its Marshall Space Flight Center (MSFC), with industry partner Ball Aerospace (contract NNM15AA18C). The Italian contribution is supported by the Italian Space Agency (Agenzia Spaziale Italiana, ASI) through contract ASI-OHBI-2022-13-I.0, agreements ASI-INAF-2022-19-HH.0 and ASI-INFN-2017.13-H0, and its Space Science Data Center (SSDC) with agreements ASI-INAF-2022-14-HH.0 and ASI-INFN 2021-43-HH.0, and by the Istituto Nazionale di Astrofisica (INAF) and the Istituto Nazionale di Fisica Nucleare (INFN) in Italy.  This research used data products provided by the IXPE Team (MSFC, SSDC, INAF, and INFN) and distributed with additional software tools by the High-Energy Astrophysics Science Archive Research Center (HEASARC), at NASA Goddard Space Flight Center (GSFC).

AV acknowledges support from the Academy of Finland grant 355672. Nordita is supported in part by NordForsk. The French contribution is supported by the French Space Agency (Centre National d'Etudes Spatiales, CNES) and the Action Thématique Processus Extrême et Multimessager of the French CNRS. JS thank GACR project 21-06825X for the support and institutional support from RVO:67985815. FC  acknowledge financial support by the Istituto Nazionale di Astrofisica (INAF) grant 1.05.23.05.06: ``Spin and Geometry in accreting X-ray binaries: The first multi frequency spectro-polarimetric campaign''. AT and SF acknowledge financial support by the Istituto Nazionale di Astrofisica (INAF) grant 1.05.24.02.04: ``A multi frequency spectro-polarimetric campaign to explore spin and geometry in Low Mass X-ray Binaries''. VEG acknowledges funding under NASA contract 80NSSC24K1403. J.Pod. acknowledges institutional support from RVO:67985815. The USRA coauthors gratefully acknowledge NASA funding through contract 80NSSC24M0035. FMV acknowledges support from the the European Union’s Horizon Europe research and innovation programme with the Marie Sk\l{}odowska-Curie (grant agreement No. 101149685). B.D.M. acknowledges support via a Ram\'on y Cajal Fellowship (RYC2018-025950-I), the Spanish MINECO grants PID2023-148661NB-I00, PID2022-136828NB-C44, and the AGAUR/Generalitat de Catalunya grant SGR-386/2021
\end{acknowledgments}

\bibliography{sample701}{}
\bibliographystyle{aasjournalv7}



\end{document}